\documentclass[twocolumn,showpacs,amsmath,amssymb,prl]{revtex4}

\usepackage{graphicx}
\usepackage{bm}

\begin{document}

\title{Electronic structure of edge and vortex states in chiral
mesoscopic superconductor.}

\author{M.\,A.\,Silaev}

\address{Institute for Physics of Microstructures, Russian Academy of Sciences,
603950, Nizhny Novgorod, GSP-105, Russia}

\date{\today}

\begin{abstract}
 We study a subgap quasiparticle spectrum in a mesoscopic disk of chiral superconductor.
 We find an exact expression for the spectrum of surface states localized at the disk edge.
  Considering an Abrikosov vortex placed at the center of a superconducting disk
 we investigate the spectrum transformation near the intersection
 points of surface and vortex anomalous energy branches .
The resulting splitting of the anomalous branches due to the
hybridization of edge and vortex states is determined by an
external magnetic field and can lead in particular to the
formation of a set of minigaps in the quasiparticle spectrum.
Tuning the external magnetic field makes it possible to control
the width of minigaps and the positions of corresponding density
of states singularities at the minigaps edges.
\end{abstract}


\maketitle

 {\bf 1. Introduction.}
 Recently, a considerable attention has been devoted to the
 investigation of a chiral superconducting state which is proposed to realize
 in p-wave triplet superconductor $Sr_2RuO_4$ (see Ref. \cite{Pwave} and references therein).
 The chiral superconductivity can be associated with a formation of
 Cooper pairs with a non-zero
 orbital angular momentum.
 In this case a value of chirality $\chi$ is determined as a projection of
 the angular momentum of Cooper pairs on the $z$
 axis. Generally, the superconducting order parameter is triplet
(singlet) for odd (even) chirality $\chi$  and is given by
$\hat\Delta=\Delta_0 e^{i\chi\theta_p}\check\sigma_z$  and
$\hat\Delta=\Delta_0 e^{i\chi\theta_p}$
correspondingly\cite{Volovik}, where $\Delta_0$ is a bulk
superconducting gap and $\check\sigma_z$ is a conventional spin
operator.
 For $\chi\neq 0$ the phase of the order parameter depends on the direction
 of the electron momentum in $xy$ plane: ${\bf p}=p
(\cos\theta_p,\sin\theta_p)$.
  The important consequence of this fact
 is an existence of surface Andreev bound states\cite{Hu,Tanaka}.
 They appear in the vicinity of scattering interfaces
 between a superconductor and an insulator, if the order parameter phase takes different values for the
 incident and reflected quasiparticles (QP) with different momentum directions.
 The formation of Andreev bound states increases
 the local density of states at the surface of a superconductor resulting in zero-bias a conductance peak anomaly
 observed in tunneling spectroscopy of high-$T_c$ cuprates with a d-wave symmetry of
 superconducting pairing \cite{Guy} as well as in p-wave triplet superconductor
 $Sr_2RuO_4$ \cite{Fogel}.

  Under an applied magnetic
 field, generating screening current and vortices the spectrum of surface states acquires a Doppler shift,
 leading to the splitting of the zero-bias conductance peak \cite{Iniotakis1}.
 Recently in work \cite{Iniotakis2} it was proposed that
 the Doppler shift effect should lead to the chirality selective
 influence of the magnetic field on the surface states in chiral p-wave
 superconductor ($|\chi|=1$), such as $Sr_2RuO_4$. The QP density of states (DOS) near the flat surface was shown
 to depend on the orientation of the magnetic field with respect to the $z$ axis as well as on the
 vorticity in case when an Abrikosov vortex is pinned near the surface
 of a superconductor.

If an Abrikosov vortex is situated close to the boundary of a
superconductor, then in addition to the Doppler shift effect
\cite{Iniotakis2} it is necessary to take into account the
hybridization of edge modes and low--energy vortex core states
\cite{CdGM}. In this case spectrum modification is determined by
the overlapping of QP wave functions localized near the surface
and near the vortex core. The characteristic localization length
of subgap QP wave functions is determined by a superconducting
coherence length $\xi$. Therefore the hybridization of vortex and
edge states should be particularly important in mesoscopic
superconducting samples of the size of several $\xi$.

Let us consider a model problem when a superconducting sample has
ideal disk geometry in $xy$ plane.
 In this case the spectrum of edge states
  can be expressed in terms of the angular
momentum $\mu$ which is conserved due to the axial symmetry. We
assume the validity of a quasiclassical approach so that one can
consider QP motion along trajectories, i.e. straight lines along
the direction of QP momentum ${\bf p}=p
(\cos\theta_p,\sin\theta_p)$.
 Employing analogy with a point Josephson junction
 the spectrum of surface states can be written as follows:
 $ E_s=-\Delta_0 \cos(\Delta\varphi/2)$,
 where $0<\Delta\varphi<2\pi$ is the
 difference between gap function phases seen by the incident
 and reflected QPs. Under the reflection of a trajectory at the disk boundary
 the angle $\theta_p$ transforms as $\theta_p\rightarrow \theta_p+\pi+2\arcsin (b/R)$,
 where $b=-\mu/k_F$ is a continuous impact parameter, i.e.
 a distance from the trajectory to the disk center, $k_F$ is a Fermi wave number and $R$ is a disk radius.
 In case when there is no circulating superconducting currents we
 obtain $\Delta\varphi=\left[\chi(\pi+2\arcsin (b/R))\right] \mod (2\pi)$ yielding a set of anomalous energy
 branches corresponding to the edge  states \cite{VolovikEdge, Stone}:
\begin{equation}\label{SurfaceBr}
 E_{sj}(\mu)\approx-(\mu-\mu_j)\omega_{sj},
 \end{equation}
 where $j=1...|\chi|$ and $\mu_j=(k_FR)\sin(\pi n_j/2\chi)$.
The integer index $n_j$ from the interval $-|\chi|< n_j<|\chi|$ is
chosen so that the combination $\chi-n_j$ to be odd. As noted in
Ref.\cite{VolovikEdge} the spectrum of edge states
(\ref{SurfaceBr}) is analogous to the general spectrum of
quasiparticles localized within a vortex core
\cite{multi-spectrum-num}. The interlevel spacing for a particular
anomalous branch $\omega_{sj}=\chi\Delta_0/(k_F R \cos(\pi
n_j/2\chi))$ is much smaller than the bulk superconducting gap
$\Delta_0$ provided $k_FR\gg 1$, therefore the anomalous branches
can be considered as functions of a continuous impact parameter
$b=-\mu/k_F$. In case of even chirality $\chi$ all energy branches
cross the Fermi level at finite impact parameters $b=-\mu_j/k_F$,
and for the odd $\chi$ there exists an energy branch with
$\mu_j=0$, crossing the Fermi level at $b=0$.

If an Abrikosov vortex is placed at the center of a
superconducting disk there appears another anomalous energy branch
associated with the spectrum of vortex states \cite{CdGM}:
 \begin{equation}\label{CdGMeq}
  E_{v}(\mu)=-\mu \omega_v.
 \end{equation}
Here $\omega_v\sim\eta\Delta_0/(k_F\xi)$, where $\xi=\hbar
V_F/\Delta_0$ is a superconducting coherence length and $V_F=\hbar
k_F/m$ is a Fermi velocity. The value of vorticity $\eta=\pm 1$ is
determined by a sign of the superfluid velocity circulation in the
counterclockwise direction around the vortex core. The angular
momentum $\mu$ in Eq.(\ref{CdGMeq}) is integer (half--integer) for
the odd (even) chirality value \cite{Volovik}.

\begin{figure}[hbt]
\centerline{\includegraphics[width=1.0\linewidth]{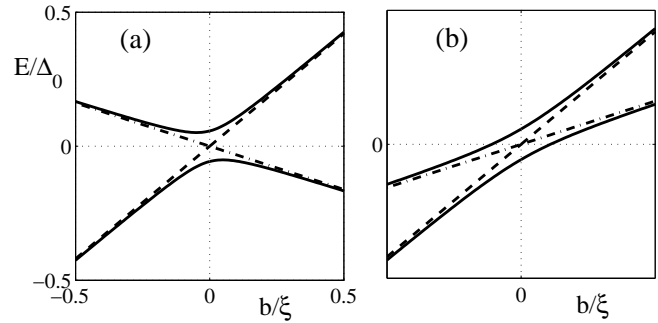}}
\caption{\label{spectrum1} FIG. 1 Shown by solid lines is the
spectrum transformation caused by the interaction of vortex and
surface spectrum branches:
 (a) chirality and vorticity have different signs ($\eta=1$,
$\chi=-1$); (b) chirality and vorticity have the same signs
($\eta=1$, $\chi=1$). The dash and dash-dotted curves correspond
to the non--interacting vortex and surface energy branches. }
\end{figure}

 Considered as continuous functions of a quasiclassical impact parameter
$b=-\mu/k_F$ the spectrum branches $E_{v}(b)$ and $E_{sj}(b)$
intersect at the certain points $b=b_j$.
 The splitting of energy levels at the degeneracy point occurs due to the hybridization
 of vortex and edge states and can be
estimated using the perturbation method for an almost degenerate
two-level system (see Ref.\cite{LL}), which yields the secular
equation:
 \begin{equation}\label{secular}
 [E-E_{sj}(b)][E-E_v(b)]=J^2,
 \end{equation}
  where the factor $J$ is determined by the
  overlapping of the corresponding wave functions.
 Using a Taylor expansion $E_v(b)=E_v(b_j)+E^\prime_v(b_j)(b-b_j)$
 and $E_{sj}(b)=E_{sj}(b_j)+E^\prime_{sj}(b_j)(b-b_j)$
   one can see that the scenario of branch splitting depends on the
  slopes of energy branches $E^{\prime}_v =d E_v/db$ and $E^{\prime}_{sj} =d E_{sj}/db$
  at the intersection point $b=b_j$.
   In case when the signs of the slopes are opposite
there appears a minigap in the QP spectrum. The minigap width i.e.
the minimal energy spacing between QP levels corresponding to the
different energy branches can be found from Eq.(\ref{secular}) as
follows:
\begin{equation}\label{mini}
 \delta E=2|J|\;\sqrt{|E^{\prime}_v E^{\prime}_{sj}|}/\left(|E^{\prime}_v|+|E^{\prime}_{sj}|\right).
\end{equation}
Otherwise, when $E^{\prime}_v$ and $E^{\prime}_{sj}$ have the same
sign,the minigap width is always equal to zero. Generally the
energy branches $E_v(b)$ and $E_{sj}(b)$ cross at $|b_j|\sim \xi$
and $|E|\sim\Delta_0$, where the spectra are not described by the
expressions (\ref{SurfaceBr},\ref{CdGMeq}). For the sake of
simplicity further we focus on a particular case of chiral p--wave
superconductor with $|\chi|=1$. Then there is only one anomalous
surface energy branch $E_s(b)$, crossing the Fermi level at $b=0$
simultaneously with the vortex energy branch $E_v(b)$. Therefore
the spectrum transformation takes place within a domain of small
energies $|E|\ll\Delta_0$, where the surface and vortex states are
well localized and the overlapping factor can be evaluated as
follows $|J|\sim \Delta_0 e^{-R/\xi}$. In this case the splitting
of energy branches is shown schematically in Fig.(\ref{spectrum1})
for the opposite and equal signs of chirality $\chi$ and vorticity
$\eta$.


 Following Ref. \cite{Iniotakis2} one could expect the
 surface energy branch $E_{s}(b)$ to be modified due to a Doppler
shift induced by a superfluid velocity circulating around the
vortex core. However, it is not the case if $|\chi|=1$ when the
Doppler shift is totally compensated by an additional
vortex--induced difference of the order parameter phase for the
incident and reflected QP. Indeed, in the presence of a
vortex--induced superfluid velocity ${\bf v_s}$ the Doppler
shifted spectrum of surface states is given by: $E_s=-\Delta_0
\cos(\Delta\varphi/2)+{\bf p_F}\cdot{\bf v_s}$, where ${\bf
p_F}\cdot{\bf v_s}=-\eta \Delta_0(b/R)$ is a Doppler shift energy
at the disk edge and the phase difference is $\Delta\varphi\approx
\pi+2(\chi+\eta)b/R$. It is easy to see that for $|\chi|=1$ the
surface energy branch is still given by Eq.(\ref{SurfaceBr}) with
$n_j=0$.


{\bf 2. Model.}
 In order to investigate in detail the effects described above we proceed with a quantitative analysis of the
  QP spectrum in chiral superconductor
on the basis of Bogoulubov - de Gennes theory:
  \begin{equation}
\hat H_0\Psi +\left(0\quad \;\,\hat \Delta\atop \hat \Delta^+\quad
0\right)\Psi =E\Psi \ ,
\end{equation}
where $\hat H_0=\check\tau_3[\left(i\hbar\nabla+ e\check\tau_3{\bf
A}/c\right)^2-p_F^2 ]/2m$, $\Psi=(u, v)$, $u$ and $v$ are the
amplitudes of the electron and hole components,
 $\check\tau_i$ are the Pauli matrices in a particle--hole space,
 $\hat\Delta$ is a gap operator:
\begin{equation}\label{gap}
  \hat\Delta=\left\{\Delta
 (\hat{\bf{r}}),e^{i\chi\theta_p}\right\},
\end{equation}
  where $\hat{\bf{r}}$ is a coordinate operator,
 $\Delta ({\bf{r}})$ describes the spatial dependence of the
 gap function and $\{A,B\}=AB+BA$ is an anticommutator which
 provides the gauge invariance of $\hat\Delta$.
 Here we omit the spin--dependent part of the gap operator $\hat
\Delta$, neglecting the spin--orbit interaction. Also we neglect
the dispersion of QP energy in the direction perpendicular to the
anisotropy plane $xy$, assuming a cylindrical Fermi surface.  The
magnetic field is directed along the $z$ axis ${\bf H}=-H{\bf
z}_0$ and for extreme type-II superconductors we can consider the
magnetic field to be homogeneous on the spatial scale $R$ and take
the gauge ${\bf A}=[{\bf H},{\bf r}]/2$. Within the quasiclassical
approach the wave function in the momentum representation can be
taken in the form:
 \begin{equation}
 \label{st}
 \Psi({\bf p})=
 \frac{2\pi}{k_F}\int\limits_{-\infty}^{+\infty} ds
 e^{-i(|{\bf p}|-\hbar k_F)s/\hbar}
\psi(s,\theta_p) \ ,
\end{equation}
where $s$ is a coordinate along a trajectory. In $(s,\theta_p)$
representation the expression for the coordinate operator in
Eq.(\ref{gap}) has the following form:
\begin{equation}\label{coordinate}
\hat {\bf r} =s {\bf k}_F/k_F
 +
 \left\{ [{\bf k}_F, {\bf z}_0], \hat\mu\right\}/(2k_F^2),
\end{equation}
 where $\hat\mu=-i\partial/\partial \theta_p$ is an angular momentum operator.
  The equation for
$\psi(s,\theta_p)$ along a quasiclassical trajectory reads:
\begin{equation}\label{QuasiBdG}
-i \check\tau_3\hbar V_F\frac{\partial}{\partial s}\psi+
\left(0\quad \;\,\hat \Delta\atop \hat \Delta^+ \quad 0\right)\psi
=\left(E+\hat\mu\,\frac{\hbar\omega_H}{2}\right)\psi,
\end{equation}
where  $\omega_H=|e|H/mc$ is a cyclotron frequency. The terms
quadratic in $H$ were neglected in (\ref{QuasiBdG}) because we
consider the distances much smaller than the cyclotron radius
$r_H=V_F/\omega_H$. The wave function in the real space is
expressed from
 Eq.(\ref{st}) in the following way (see Refs.~\cite{PRB2007},\cite{MS2006}):
 \begin{equation}\label{xy}
  \Psi(r,\theta)=\int_0^{2\pi}d\theta_pe^{ik_F r\cos(\theta-\theta_p)}
  \psi(r\cos(\theta-\theta_p),\theta_p).
 \end{equation}
   For an ideal disk it is convenient to use a
   polar coordinate system $(r,\theta)$ with the origin
 at the disk center. Then, the
boundary condition at the surface of a superconducting disk of the
radius $R$ reads:
 \begin{equation}\label{bcGen}
  \Psi (R,\theta)=0.
\end{equation}


{\bf 3. Spectrum of edge states.}
 At first we investigate the spectrum of edge states solving Eq.(\ref{QuasiBdG})
  with spatially homogeneous order parameter distribution: $\Delta({\bf r})=\Delta_0$
 and applying the boundary condition (\ref{bcGen}).
  Due to the axial symmetry of the superconducting sample
 we separate the $\theta_p$ and $s$ variables in Eq.(\ref{QuasiBdG}):
\begin{equation}\label{AngMom}
  \psi (s,\theta_p)= e^{i\mu\theta_p+i\chi\check\tau_3\theta_p/2}G_{\mu} (s),
\end{equation}
    where $\mu=n+\chi/2$ is an angular momentum, and $n$ is integer. The function
 $G_{\mu}$ satisfies the following equation:
 \begin{equation}\label{g1}
 -i\check\tau_3\hbar V_F\frac{\partial}{\partial s}
 G_{\mu}+\Delta_0\check\tau_1 G_{\mu}=
 \tilde{E} G_{\mu},
 \end{equation}
where $\tilde{E}=E+\mu\,(\hbar\omega_H/2)$. In order to apply the
boundary conditions (\ref{bcGen}) we evaluate the integral in
(\ref{xy}) using the stationary phase method. For a given $\mu$ we
obtain:
 $$
 \Psi (R,\theta)= e^{i(k_Fs^*-\pi/4)}\psi (s^*,\theta_{1})+e^{i(\pi/4-k_Fs^*)}\psi (-s^*,\theta_{2}),
 $$
 where $s^*=\sqrt{R^2-(\mu/k_F)^2}$ and
 the stationary phase points are given by
 $\theta_{1}=\theta+\arcsin (\mu/k_F R)$
 and $\theta_{2}=\theta+\pi-\arcsin (\mu/k_F R)$.
 Thus, we obtain the boundary condition for the function $G_\mu
 (s)$:
\begin{equation}\label{bc1}
G_\mu (s^*)=e^{i\alpha-i\check\tau_3\varphi}G_\mu(-s^*),
\end{equation}
where $\alpha=\mu[\pi-2\arcsin(\mu/k_FR)]-2k_Fs^*-\pi/2$ and
$\varphi=\chi[\arcsin(\mu/k_FR)-\pi/2]$.

 The general solution of Eq.(\ref{g1})
 can be written as follows:
 \begin{equation}\label{solution1}
 G_\mu=c\left(1 \atop e^{i\gamma}\right) e^{qs/\xi}+d\left(1 \atop e^{-i\gamma}\right)
 e^{-qs/\xi},
 \end{equation}
 where $c, d$ are the scalar coefficients,
 $\gamma=\arccos(\tilde{E}/\Delta_0)$ and
 $q=\sqrt{\Delta_0^2-\tilde{E}^2}/\Delta_0$.
  Then, from the boundary condition (\ref{bc1}) we obtain the expression
 for the spectrum of edge states:
 \begin{equation}\label{spSurfGen}
E=\Delta_0M/\sqrt{1+M^2}-\mu (\hbar\omega_H/2),
\end{equation}
 where
  $
 M=\coth(2qs^*/\xi) \cot\varphi-\cos\alpha /\sinh (2qs^*/\xi).
 $
 It is possible to evaluate Eq.(\ref{spSurfGen}) to find an explicit expression for the energy levels
 lying much lower than the superconducting
 gap. For simplicity we start our analysis with the case of a zero
 magnetic field. Considering the low energies $|E|\ll\Delta_0$
 we obtain that the spectrum consists of $|\chi|$ energy branches:
 \begin{equation}\label{spSurf1}
 E_{sj}(\mu)=-(\mu-\mu_j)\omega_{sj}+\Delta_0\frac{(-1)^{k_j}\cos\alpha}{\sinh(2s^*/\xi)},
 \end{equation}
 where $\omega_{sj}=\chi\Delta_0\coth(2s^*/\xi)/(k_F R\cos(\pi
 n_j/2\chi))$ and $\mu_j=(k_FR)\sin(\pi n_j/2\chi)$. The integer index $n_j$
 from the interval $-|\chi|< n_j<|\chi|$ is chosen so that the combination $\chi-n_j$
 to be odd: $\chi-n_j=2k_j+1$.
 From Eq.(\ref{spSurf1}) one can see that the
 energy levels are the oscillating functions of a disk radius with a period $\delta
 R=\pi/k_F$.
The amplitude of energy levels oscillations is larger than the
interlevel spacing provided the disk radius is smaller than the
critical value $R_c$ determined by the condition
$\omega_{sj}\sim\Delta_0/\sinh (2s^*/\xi)$. For the typical values
of the parameter $k_F\xi\sim 10^2-10^3$ we obtain $R_c/\xi \sim
3-5$.
 Note that in case $|\chi|= 1$  Eq.(\ref{spSurf1}) is completely
analogous to the expression obtained in Ref.\cite{MelnikovPrl} for
the spectrum of vortex core states modified by the normal
reflection of QP at the surface of s-wave mesoscopic
superconductor. At $R\gg R_c$ the exponentially small oscillating
term in Eq.(\ref{spSurf1}) can be omitted and we obtain the
low--energy spectrum of surface states consistsing of a set of
anomalous branches (\ref{SurfaceBr}), similar to the spectrum of
multiquantum vortex with a vorticity equal to $\chi$
\cite{multi-spectrum-num}. Then, Eq.(\ref{spSurf1}) describes an
appearance of an energy band due to the interaction of surface
states localized at the opposite ends of a trajectory $s=\pm s^*$.
The bandwidth $\Delta_0/\sinh(2s^*/\xi)$ is proportional to the
overlap of decaying wave functions of edge states.

\begin{figure}[hbt]
\centerline{\includegraphics[width=0.7\linewidth]{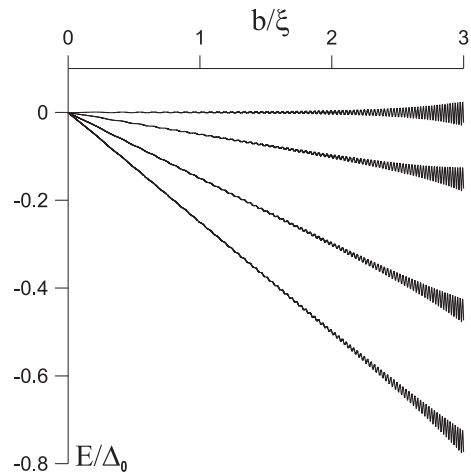}}
\caption{\label{plotS1}FIG. 1 Spectrum of surface states for
$\chi=-1$ and $\mu=2n+1/2$. Curves from bottom to top correspond
to the different values of magnetic field from $H=0$ to
$H=2\phi_0/(\xi R)$ (corresponding to $\omega_s=0$). We choose
$R=4\xi$ and $k_F\xi=200$.}
\end{figure}

 Applying a magnetic field $H$ along the $z$ axis one introduces
 the shift of surface energy levels $\mu(\hbar\omega_H/2)$
 in Eq.(\ref{spSurf1}). In fact it is a Doppler shift effect due to the Meissner
 current flowing along the circumference of
  a superconducting disk. The same effect was studied in
  Ref.\cite{Iniotakis2} for the flat geometry of the superconducting sample boundary.
 Let us consider the expression (\ref{spSurf1}) in more detail for the case of
 $|\chi|=1$.
 Taking into account the finite external magnetic
 field we obtain that the spectrum is given by Eq.(\ref{spSurf1})
 with $\omega_{s1}=\omega_{s}=\chi\Delta_0/(k_F R)+ \hbar\omega_H/2$ and
 the phase of energy oscillations: $\alpha(\mu)=\mu\pi-2k_FR-\pi/2 $.
 Note that $\alpha(\mu+2)=\alpha(\mu) +2\pi$, therefore the
 spacing between levels corresponding to the angular momentum values $\mu$ and $\mu+2$ is
 $2\omega_s$ and can be neglected within the quasiclassical consideration. On the other hand, the
 spacing between levels corresponding to $\mu$ and $\mu+1$
 is $\Delta_0 e^{-2R/\xi}\sin(2k_FR)$, which can be much larger
 than $\omega_s$ if $R<R_c$. Thus one can consider
 two continuous branches corresponding to the odd and even values of
 $n=\mu-1/2$:
 \begin{equation}\label{spSurf}
  E_s(b)=\omega_sk_F b  \pm
  2\chi\Delta_0 e^{-2R/\xi}\sin(2k_FR),
 \end{equation}
 where $b=-\mu/k_F$ is an impact parameter.
  At  $R>\xi$, the Doppler shift energy $k_Fb(\hbar\omega_H/2)$ in Eq.(\ref{spSurf}) can
 substantially change the slope of the branches $E_s(b)$.
     Indeed, $\hbar \omega_H\sim (H/H_{c2})\Delta_0/(k_F \xi )$, where $H_{c2}\sim \phi_0/\xi^2$
   is the upper critical field and $\phi_0=\pi\hbar c/e$ is the magnetic flux quantum. Therefore the magnetic field of
   the magnitude $|H|>(\xi/R)H_{c2}$ can reverse the
   slope of $E_s(b)$. Particularly at $H=-2\chi\phi_0/(\xi R)$
   we obtain a dispersionless energy branches $E_s=\pm
   2\Delta_0e^{-2R/\xi}\cos(2k_FR)$. In the Fig.(\ref{plotS1}) we show one of the spectrum branches
   (for $\mu=2n+1/2$) given by Eq.(\ref{spSurfGen}) for the
   different values of magnetic field.
   Considering the measurable characteristics of the QP
   spectrum we obtain that the density of states (DOS) at the Fermi level
 $\nu (0)=1/|\omega_s|$ can be controlled tuning the magnitude and
 direction of an external magnetic field.

Analyzing the influence of external magnetic field on the surface
states spectrum we have neglected a magnetic field $H_s$ generated
by the current, carried by the surface states. The density of this
current is of the order of the critical one for depairing
\cite{VolovikEdge,Stone} and it flows within the surface shell of
the width $\xi$. Evaluating the magnetic field we obtain $H_s\sim
(\xi/\lambda)^2 H_{c2}$, where $\lambda$ is a London penetration
length. Thus for extreme type--II superconductors ($\xi/\lambda\ll
1$) the field of the surface current can be neglected.

   The considered model with a spatially
    homogeneous gap function $\Delta ({\bf r})=\Delta_0$
 is adequate only for not very large applied magnetic field. Generally, it
 does not work when the field is large enough to suppress a
  surface barrier preventing vortex entry $H_c\sim
  H_{c2}(\xi/R)$ \cite{Buzdin}. Certainly, in our case a criterion for the
  vortex formation should be sensitive to the orientation of magnetic
  field with respect to the $z$ axis. Now we proceed with the analysis of the
  QP spectrum assuming that a vortex has already
  entered the sample and is placed at the center of a superconducting disk.

  {\bf 4. Vortex--induced spectrum transformation.}
   In the vicinity of a vortex core the gap function has the following form:
    $\Delta ({\bf r})=\Delta_0 D_v(r) e^{i\eta\theta},$
 where $\eta=\pm 1$ is the vorticity and $D_v(r)$ is a dimensionless
 vortex core profile. The gap operator $\hat \Delta$ in $(s,\theta_p)$ representation
 has the following form:
\begin{equation}\label{QuasiD}
 \hat \Delta =\Delta_0\frac{D_v(s)}{2|s|}\left\{(s+\eta\hat\mu/k_F),e^{i(\chi+\eta)\theta_p}\right\}.
\end{equation}
 To be specific, we choose a model vortex core profile:
 $D_v(r)=r/\sqrt{r^2+\xi^2}$.
Once again we can separate the $s$ and $\theta_p$ variables:
 \begin{equation}\label{AngMom1}
  \psi (s,\theta_p)=e^{i\mu\theta_p+i(\chi+\eta)\check\tau_3\theta_p/2}G_{\mu} (s),
\end{equation}
 where $\mu$ is integer. We will consider the trajectories passing
 close to the vortex core with impact parameters $b\ll\xi$, corresponding to $|\mu|\ll
 k_F\xi$. Then, from the boundary condition (\ref{bcGen}) we obtain Eq.(\ref{bc1})
 with $s^*=R$, $\alpha=\mu\pi-2k_FR-\pi/2$ and
 $\varphi=(\chi+\eta )[\mu/(k_FR)-\pi/2]$.
 The function $G_{\mu}(s)$ we satisfies the following equation:

 \begin{equation}\label{g2}
  -i\hbar V_F\check\tau_3\frac{\partial}{\partial s}
 G_{\mu}+{\bf U}(s) G_{\mu}=
  \left(\tilde{E}-\frac{\mu}{k_F\xi}{\bf W}(s)\right) G_{\mu}.
 \end{equation}
 The matrices ${\bf U}, {\bf W}(s)$ are defined at $-R<s<R$ as follows:
 $$
 {\bf U}(s)=\Delta_0 D_v(s) \frac{s}{|s|}\check\tau_1
 $$
 $$
 {\bf W}(s)=\Delta_0D_v(s)\frac{\xi}{|s|}\check\tau_2
 $$
We assume that the size of the disk is rather large: $R>\xi$
therefore Eq.(\ref{g2}) together with the boundary condition
(\ref{bc1}) describe the interaction of vortex states localized
near the vortex center i.e. $s=0$ and the edge states localized at
$s=\pm R$. Thus, it is natural to use the tight--binding
approximation of the wave function to calculate the spectrum. To
this end we consider a continuation of the function $G_{\mu}(s)$
and coefficients ${\bf U}, {\bf W}(s)$ of Eq.(\ref{g2}) to the
whole axis $-\infty<s<\infty$ using the periodicity conditions:
 $$
 G_\mu(s+2R)=e^{i\alpha-i\check\tau_3\varphi}G_{\mu}(s)
 $$
 $$
 {\bf U}, {\bf W}(s+2R) = e^{-2i\check\tau_3\varphi}{\bf U}, {\bf
 W}(s).
 $$

 We find the solution of Eq.(\ref{g2}) as a
 superposition of the functions localized at $s_n=2nR$ corresponding
 to the vortex states and localized at $d_n=(2n+1)R$ corresponding
 to the surface states:
\begin{equation}\label{solTB2}
G_\mu=C_1\sum_n V_n+C_2\sum_n S_n,
\end{equation}
 where $C_1, C_2$ are the arbitrary coefficients and the generic terms are:
 $$
 V_n (s)=e^{in(\alpha-\check\tau_3\varphi)}e^{-K_v(s-s_n)}e^{i\check\tau_3\pi/4}\left(1\atop 1\right),
 $$
 \begin{equation}
 \nonumber
 K_v(s-s_n)=\int_{s_n}^s D_v(s-s_n)
 \vartheta(s-s_n) \frac{ds}{\xi}
\end{equation}
 and
 $$
 S_n (s)=e^{in(\alpha-\check\tau_3\varphi)}
 e^{-K_s(s-d_n)}e^{-i\check\tau_3\pi/4}\left(1\atop 1\right),
 $$
  \begin{equation}
   \nonumber
 K_s(s-s_n)=\int_{d_n}^s D_v(s-d_n) \vartheta(s-d_n)
 \frac{ds}{\xi}
 \end{equation}
where we have introduced the step function $\vartheta(s)=s/|s|$.

Following the standard tight binding method we substitute the
solution in the form (\ref{solTB2}) into the Eq.(\ref{g2}),
 multiply by $ V^*_n$ and $ S^*_n$ from the left and
 integrate over $s$, taking into account the overlapping of the nearest
 neighbor functions. We omit here the details of calculation of the corresponding integrals
 which yields a linear system of equations for the coefficients
 $C_1,C_2$:
 \begin{eqnarray}\label{systemC}
 \nonumber
  \left[E-E_v(\mu)\right]C_1
  =-(i/2)e^{-R/\xi}(1-e^{-i\alpha})C_2\\
   \left[E-E_s(\mu)\right]C_2=ie^{-R/\xi}(1-e^{i\alpha})C_1.
 \end{eqnarray}
The solvability condition for this system is given by the
Eq.(\ref{secular}) with
$J=\sqrt{2}\Delta_0e^{-R/\xi}\sin(\alpha/2)$.
 The vortex energy branch $E_v(\mu)$
is given by (\ref{CdGMeq}) with $\omega_v\approx
0.84\eta\Delta_0/(k_F\xi)+\hbar\omega_H/2$ and the surface energy
branch $E_s(\mu)$ is given by (\ref{SurfaceBr}) with $n_j=0$ and
  $\omega_s\approx \chi\Delta_0/(k_F R)+\hbar\omega_H/2$.

 According to the arguments presented in introduction
  in case when the vorticity $\eta$ and chirality $\chi$ are of
the opposite signs [Fig.(\ref{spectrum1})a], there exists a
minigap at the Fermi level. In this case the DOS has van Hove
singularities at the minigap edges, i.e. at $E=\pm \delta E$.
Applying the magnetic field $H$ along the $z$ axis one can change
the slope of the intersecting branches $E_v(b)$ and $E_s(b)$ and
therefore shift the positions of van Hove singularities according
to the Eq.(\ref{mini}). For large enough magnetic fields
$|H|>2\phi_0/(\xi R)$ the slope of energy branch $E_s(b)$ can be
reversed. In this case as well as in case of the equal signs of
vorticity $\eta$ and chirality $\chi$ the spectrum is gapless as
shown on the [Fig.(\ref{spectrum1})b].


{\bf 5. Effect of surface roughness.}
 In conclusion we note that in previous sections we analyzed the
 electronic spectrum for a sample with the perfect surface. One can expect
 that the surface roughness would break the interference of QP waves
 and as a result the DOS singularity at the minigap edge would be smeared.
 However, it is not the case for a wide class of surface
 imperfections. Particularly, let us assume that the surface of the sample is described by the
 equation $r=R(\theta)$, where $R(\theta)$ is a smooth function
 fluctuating over an average value $r=R_0$, so that $|R(\theta)-R_0|\ll\xi$
 and $|dR/d\theta|\ll\xi$.  Then, following a procedure developed in Ref.\cite{PRB2007}
one can find the solution of Eq.(\ref{QuasiBdG}) in the form of
expansion (\ref{solTB2}) with the angle--dependent coefficients
$C_1, C_2 (\theta)$. Using a tight binding method we arrive at the
system of differential equations for the functions
$C_1,C_2(\theta)$, which coincides with Eq.(\ref{systemC})
 if one takes $\hat\mu=-i\partial/\partial \theta$ and
 $e^{i\alpha}=-ie^{-ik_F(R(\theta)+R(\theta+\pi))}e^{i\pi\hat\mu}$.
  Considering the low--energy limit $|E|\ll\Delta_0$,
 it is natural to assume that the functions $C_{1,2}(\theta)$
  consist of a limiting number of the lowest angular harmonics with $|\mu|\ll k_F\xi$.
 Thus, the coefficients of Eq.(\ref{systemC}) can be averaged over the small
 angular interval to exclude higher angular harmonics of the
 rapidly oscillating exponent $e^{-ik_F(R(\theta)+R(\theta+\pi))}$.
 It is important that the right hand sides of equations in system
 (\ref{systemC}) can not vanish after this averaging, in contrast
 to the analogous problem for the case of s--wave
 superconductor\cite{PRB2007}. The resulting spectrum
 should have the form
  (\ref{secular}) with $J=\beta\Delta_0e^{-R_0/\xi}$,
 where a factor $\beta\sim 1$ depends on the particular
 realization of the surface roughness.

 I am grateful to G.E. Volovik for drawing my
attention to this problem and to A.S. Melnikov for numerous
stimulating discussions and help with preparation of this paper.
Also it is my pleasure to thank E. Ezhova for help with numerical
calculations and A. Aladyshkin for valuable remarks. The research
was supported, in part, by Russian Foundation for Basic Research,
by Program ``Quantum Macrophysics'' of RAS, and by Russian Science
Support and ``Dynasty'' Foundations.

\end{document}